# Hawking Radiation as Tunnelling in Static Black Holes

Wenbiao Liu*

*Department of Physics, Institute of Theoretical Physics, Beijing Normal University, Beijing 100875, China*

## Abstract

Hawking radiation can usefully be viewed as a semi-classical tunnelling process that originates at the black hole horizon. The conservation of energy implies the effect of self-gravitation. For a static black hole, a generalized Painleve coordinate system is introduced, and Hawking radiation as tunnelling under the effect of self-gravitation is investigated. The corrected radiation is consistent with the underlying unitary theory.

## I. INTRODUCTION

The thermal Hawking radiation[1,2] implies the loss of unitary, and then the breakdown of quantum mechanics.[3] According to Hawking radiation, black hole can radiate its energy away, and vanish in the end. Where does the information go? Although Hawking radiation can be described in a unitary theory in string theory,[4] it is not clear how information is returned.

Originally, Hawking explained the existence of black hole radiation as particle's tunnelling due to vacuum fluctuations near the horizon. The radiation is like electron-positron pair creation in a constant electric field. The energy of a particle can change its sign after crossing the horizon. So a pair created by vacuum fluctuations just inside or outside the horizon can materialize with zero total energy, after one member of the pair has tunnelled to the opposite side. However, Hawking's original derivation did not proceed in this way.[1] There were two difficulties to overcome. The first was to find a well-behaved coordinate system at the event horizon. The second was where the barrier is. Recently, a method to describe Hawking radiation as tunnelling process was developed by Kraus and Wilczek[5] and elaborated by Parikh and Wilczek.[6−8] This method gives a leading correction to the emission rate arising from the loss of mass of the black hole corresponding to the energy carried by the radiated quantum. Following this method, the radiation from AdS black hole and de Sitter cosmological horizon were also studied.[9−11] All these spherically symmetric investigations are successful. In this paper, we will investigate Hawking radiation from generalized static black holes, and then give some internal characteristics of the tunnelling. After the general calculation, we can find that the tunneling from a static black hole is a quasi-equilibrium process. The energy, entropy, and temperature of the black hole system satisfy the relation $dM = TdS$, and the entropy does not disappear.

---

*E-mail: wbliu@bnu.edu.cn

The remainder of the paper is organized as follows. In Sec. 2, after a lot of static metrics are introduced, the general static black hole metric will be concluded in details, and genealized Painleve coordinates will be given. In Sec. 3, using the generalized Painleve coordinate system, we calculate the rate of emission via tunnelling process. In the end, a brief discussion will be given.

## II. STATIC BLACK HOLES AND PAINLEVE COORDINATE SYSTEM

The general static black hole metric can be expressed as

$$ds^2 = g_{00}dt^2 + g_{11}dr^2 + g_{22}d\theta^2 + g_{33}d\varphi^2, \qquad (1)$$

where $g_{00}, g_{11}, g_{22}, g_{33}$ are functions not including time coordinate $t$.

The line element of Schwarzschild black hole is

$$ds^2 = -(1 - \frac{2M}{r})dt^2 + (1 - \frac{2M}{r})^{-1}dr^2 + r^2(d\theta^2 + \sin^2\theta d\varphi^2), \tag{2}$$

where $M$ is the mass of the black hole.

The line element of Reissner-Nordstrom black hole is

$$ds^2 = -(1 - \frac{2M}{r} + \frac{Q^2}{r^2})dt^2 + (1 - \frac{2M}{r} + \frac{Q^2}{r^2})^{-1}dr^2 + r^2(d\theta^2 + \sin^2\theta d\varphi^2), \tag{3}$$

where $M, Q$ are the mass and static charge of the black hole respectively.

The Garfinkle-Horowitz-Strominger(GHS) dilatonic black hole can be expressed as[12,13]

$$ds^2 = -(1 - \frac{2M}{r})dt^2 + (1 - \frac{2M}{r})^{-1}dr^2 + r(r-a)(d\theta^2 + \sin^2\theta d\varphi^2), \tag{4}$$

where $a = \frac{Q^2}{2M}e^{-2\phi_0}$, $\phi_0$ is the asympototic value of the dilaton field, $M$ is the mass, and $Q$ is the magnetic charge.

The metric of the static Gibbons-Maeda(GM) dilaton black hole is described by[14]

$$ds^2 = -\frac{(r-r_+)(r-r_-)}{R^2}dt^2 + \frac{R^2}{(r-r_+)(r-r_-)}dr^2 + R^2(r)(d\theta^2 + \sin^2\theta d\varphi^2), \tag{5}$$

where $r_\pm = M \pm \sqrt{M^2 + D^2 - P^2 - Q^2}$, $D = (P^2 - Q^2)/(2M)$ and $R^2 = r^2 - D^2$. The parameters $Q$ and $P$ represents electric charge and magnetic charge, respectively.

The Garfinkle-Horne(GH) dilaton black hole metric in Einstein-Maxwell dilaton theory can be expressed as[12,15]

$$ds^2 = -(1 - \frac{r_+}{r})(1 - \frac{r_-}{r})^{(1-a^2)(1+a^2)}dt^2 + (1 - \frac{r_+}{r})^{-1}(1 - \frac{r_-}{r})^{(a^2-1)(1+a^2)}dr^2$$
$$+ r^2(1 - \frac{r_-}{r})^{2a^2/(1+a^2)}(d\theta^2 + \sin^2\theta d\varphi^2), \tag{6}$$

with dilaton field $e^{2\phi} = (1 - r_-/r)^{2a/(1+a^2)}e^{2\phi_0}$ and Maxwell field $F = (Q/r^2)dt \wedge dr$, where $a$ is a coupling constant. Mass $M$ and chrage $Q$ of the black hole are related to parameters $r_+$ and $r_-$ by

$$2M = r_+ + (\frac{1-a^2}{1+a^2})r_-, \quad Q^2 = \frac{r_+ r_-}{1+a^2}e^{2a\phi_0}.$$

The line element in Schwarzschild-de Sitter spacetime can be written as

$$ds^2 = -(1 - \frac{2M}{r} - \frac{1}{3}\Lambda r^2)dt^2 + (1 - \frac{2M}{r} - \frac{1}{3}\Lambda r^2)^{-1}dr^2 + r^2(d\theta^2 + \sin^2\theta d\varphi^2), \qquad (7)$$

where $M$ is the mass of the black hole, $\Lambda > 0$ is the cosmological constant.

Thinking of the static black holes we encountered before[16,17] and several examples given as above, we simplify the line element expression Eq.(1) into the following

$$ds^2 = -f(r)dt^2 + f^{-1}(r)dr^2 + R^2(r)(d\theta^2 + \sin^2\theta d\varphi^2), \qquad (8)$$

so apparently what we will discuss next can be applied to many black holes including Schwarzschild, Reissner-Nordstrom, all kinds of Dilaton, and (anti-)de Sitter etc.

Thinking of the metric Eq.(8), we need a transformation to make none of the components of either the metric or the contra metric diverge at the horizon. Moreover, constant time slices are just flat Euclidean in radial. To obtain a coordinate system analogous to Painleve coordinates[18], we should perform a coordinate transformation

$$dt = d\tau - F(r)dr, \qquad (9)$$

where $F(r)$ is a function of $r$, independent on $t$.

Putting Eq.(9) into the line element expression (8), we have

$$ds^2 = -f(r)(d\tau - F(r)dr)^2 + f^{-1}(r)dr^2 + R^2(r)(d\theta^2 + \sin^2\theta d\varphi^2)$$
$$= -f(r)d\tau^2 + 2f(r)F(r)d\tau dr + (f^{-1}(r) - f(r)F^2(r))dr^2 + R^2(r)(d\theta^2 + \sin^2\theta d\varphi^2). \qquad (10)$$

As a corollary, we demand that the metric is flat Euclidean in radial to the constant-time slices. We then get the condition

$$f^{-1}(r) - f(r)F^2(r) = 1,$$

that is

$$F^2(r) = f^{-1}(r)[f^{-1}(r) - 1]. \qquad (11)$$

So, Eq.(10) can be changed into

$$ds^2 = -f(r)d\tau^2 + 2f(r)F(r)d\tau dr + dr^2 + R^2(r)(d\theta^2 + \sin^2\theta d\varphi^2). \tag{12}$$

There is now no singularity at the event horizon, and the true character of the spacetime, as being stationary but not static, is manifest.

For later usage, let us evaluate the radial, null geodesics described by Eq.(12) as following

$$-f(r) + 2(1 - f(r))^{1/2}\,\dot{r} + \dot{r}^2 = 0,$$

where a dot means derivation with respect to $\tau$. Solving the quadratic, we then have

$$\dot{r} = \frac{dr}{d\tau} = \pm 1 - [1 - f(r)]^{1/2}, \tag{13}$$

where the +(-) sign can be identified with outgoing (ingoing) radial motion, respectively, under the assumption that $\tau$ increases towards future.

Let us now focus on a semiclassical treatment of the associated radiation. We adopt the picture of a pair of virtual particles spontaneously created just inside the horizon. The positive energy virtual particle can tunnel out while the negative one is absorbed by the black hole resulting in a decrease in the mass. The particle is considered as a shell (an ellipsoid shell) of energy $\omega$. We fix the total ADM mass and let the mass $M$ of the static black hole vary. If a shell of energy (mass) $\omega$ is radiated outwards the outer horizon, the static black hole mass will be reduced to $M - \omega$. We should replace $M$ with $M - \omega$ in the metric Eq.(12) and the geodesic Eq.(13) to describe the moving of the shell.

### III. TUNNELLING PROCESS

We evaluate the imaginary part of the action for an outgoing positive energy particle which crosses the horizon outwards from $r_{in}$ to $r_{out}$. The imaginary part of the action is

$$\text{Im}\,I = \text{Im}\int_{r_{in}}^{r_{out}} p_r dr = \text{Im}\int_{r_{in}}^{r_{out}}\int_0^{p_r} dp' dr. \tag{14}$$

We make the transition from momentum variable to energy variable using Hamilton's equation $\dot{r} = \frac{dH}{dp_r}$. Thinking about Eq.(13) in the vicinity of event horizon, the result is

$$\operatorname{Im} I = \operatorname{Im} \int_{r_{in}}^{r_{out}} \int_{M}^{M-\omega} \frac{dM' dr}{1 - [1 - f(r)]^{1/2}}. \tag{15}$$

After some calculations and switching the order of integration, we can rewrite the above equation as following

$$\operatorname{Im} I = \operatorname{Im} \int_{M}^{M-\omega} \int_{r_{in}}^{r_{out}} \frac{1 + [1 - f(r)]^{1/2}}{f(r)} dr dM' = \operatorname{Im} \int_{M}^{M-\omega} \int_{r_{in}}^{r_{out}} \frac{1 + [1 - f(r)]^{1/2}}{f'(r)|_{r=r_+}(r - r_+)} dr dM'. \tag{16}$$

Thinking of the integration formula

$$\int_{r_1}^{r_2} \frac{g(r)}{r - a} dr = -i\pi g(a), \tag{17}$$

where $r_1 \leq a \leq r_2$, we can finish the integration of $r$ in Eq.(16) and get

$$\operatorname{Im} I = \operatorname{Im}(-i\pi) \int_{M}^{M-\omega} \frac{1 + [1 - f(r_+)]^{1/2}}{f'(r)|_{r=r_+}} dM' = -2\pi \int_{M}^{M-\omega} \frac{dM'}{f'(r)|_{r=r_+}}. \tag{18}$$

According to thermodynamics of static black holes, Hawking temperature can be calculated by

$$T = \frac{\kappa}{2\pi} = \frac{1}{2\pi}(-\frac{1}{2} \lim_{r \to r_+} f'(r)) = -\frac{1}{4\pi} f'(r)|_{r=r_+}. \tag{19}$$

Putting Eq.(19) into Eq.(18) and thinking of the first law in thermodynamics $dM = TdS$, we get

$$\operatorname{Im} I = \frac{1}{2} \int_{M}^{M-\omega} \frac{dM'}{T} = \frac{1}{2} \int_{M}^{M-\omega} dS. \tag{20}$$

Apparently the emission rate depends only on the entropy difference between before and after the radiation, and it can be expressed as

$$\Gamma = e^{-2 \operatorname{Im} I} = \exp(\Delta S). \tag{21}$$

This is consistent with an underlying unitary theory.

## IV. CONCLUSIONS AND DISCUSSIONS

After considering the self-gravitation in generalized static spherical black holes, we conclude that the emission rate can be expressed as the standard form Eq.(21). The radiation rate can be written as the exponent of the difference in the Bekenstein-Hawking entropy, $\Delta S$, before and after emission.[7]

Thinking of Eq.(20) and Eq.(21) more considerably, we can find that the radiation is really thermal as before for every quantum while the temperature can be regarded as a constant. The formula is the same as thermal spectrum $\Gamma = e^{-\beta \omega}$. But for a macro process, $T$ does not keep as a constant, the emission will not be thermal. However, the unitary theory is still satisfied during the whole process.

This is precisely consistent with an underlying unitary theory. According to Ref.[6], "Quantum mechanics tells us that the rate must be expressed as

$$\Gamma(i \to f) = |M_{fi}|^2 \cdot (phase \quad space \quad factor), \tag{22}$$

where the first term on the right is the square of the amplitude for the process. The phase space factor is obtained by summing over final states and averaging over initial states. But the number of final states is just the exponent of the final entropy, while the number of initial states is the exponent of the initial entropy." So, we have

$$\Gamma = \frac{e^{S_{final}}}{e^{S_{intial}}} = \exp(\Delta S), \tag{23}$$

which is in agreement with our result Eq.(21). This suggests that the formula we have is actually exact, up to a prefactor.

Moreover, we have investigated Hawking radiation from generalized static black holes, and then given some internal characteristics of the tunnelling. After the general calculation, we can find that the tunneling from a static black hole is a quasi-equilibrium process. The energy, entropy, and temperature of the black hole system satisfy the relation $dM = TdS$ during the whole process, and the entropy does not disappear. Our result Eq.(21) not only is correct to all the static spherical black holes from Eq.(2) to Eq.(7), or even more, but also gives the truth of the quasi-equilibrium emission process.

I would like to give great thanks to Prof. Zheng Zhao for helpful discussions. This research is supported by the National Natural Science Foundation of China (Grant No. 10373003, 10475013) and the National Basic Research Program of China (Grant No. 2003CB716302).